\documentclass[preprintnumbers,amsmath,amssymb,floatfix,11pt,prd,onecolumn,showpacs,
superscriptaddress,nofootinbib]{revtex4}
\usepackage{graphicx}
\usepackage{epsfig}
\usepackage{bm}
\usepackage{amsfonts}

\begin{document}

\title{\large{\bf ROLE OF GENERALIZED COSMIC CHAPLYGIN GAS IN ACCELERATING UNIVERSE : A FIELD THEORETICAL PRESCRIPTION}}

\author{ \textbf{Prabir
Rudra}}\email{prudra.math@gmail.com} \affiliation{Department of
Mathematics, Bengal Engineering and Science University, Shibpur,
Howrah-711 103, India.\\
Department of Mathematics, Pailan College
of Management and Technology, Bengal Pailan Park, Kolkata-700 104,
India.}

\begin{abstract}
In this paper we investigate the role played by dark energy in the
form of Generalized cosmic Chaplygin gas in an accelerating
universe described by FRW cosmology. We have tried to describe the
model from the theoretical point of view of a field, by
introducing a scalar field $\phi$ and a self interacting potential
$V(\phi)$. The corresponding expressions for the field are
obtained for the given model. Statefinder parameters have been
used to characterize the dark energy model. Plots have been
generated for characterizing different phases of universe
diagrammatically and a comparative study is performed with the
Modified Chaplygin gas model. As an outcome of the study,
Generalized cosmic Chaplygin gas is identified as a much less
constrained form of dark energy as compared to modified Chaplygin
gas.

\end{abstract}

\maketitle

\newpage

\section{INTRODUCTION}

\noindent

At the turn of the last century a new feather was added to Edwin
Hubble's discovery of the expanding universe, when it was
confirmed from the observations of type Ia supernovae and CMBR
data that the expansion is having an accelerating nature
associated with it \cite{perl, spr}. It marked the beginning of a
new era in cosmology, as it gave rise to the new concept of the
mysterious dark energy(DE) \cite{riess}. DE, as the name suggests
is an invisible form of energy, possessing a large negative
pressure component, with the help of which it is able to bring
about the cosmic acceleration. Hence this type of matter violates
the strong energy condition, i.e., $\rho+3p<0$.

Now, soon after this attribution, the entire cosmological society
began to investigate for a suitable model that can effectively
play the role of DE. The simplest form of DE was found to be a
tiny cosmological constant, $\Lambda$, which gave rise to the
simplest model of the accelerating universe, the $\Lambda$CDM
model. DE associated with a scalar field \cite{noj1} is called
quintessence, and provides an effective model of the accelerating
universe. As time rolled on various Chaplygin gas models came into
existence. Extensive research saw Chaplygin gas (CG) \cite{kam}
pave way to Generalized Chaplygin gas (GCG) \cite{gor} and
Modified Chaplygin gas (MCG) \cite{ben, deb}. The MCG equation of
state is given by
\begin{equation}
p=A\rho-\frac{B}{\rho^\alpha}
\end{equation}
where $p$ and $\rho$ are respectively the pressure and energy
density and $0\leq\alpha\leq1$, $A$ and $B$ are positive
constants.

In 2003, P. F. Gonz´alez-Diaz \cite{gon} introduced the
Generalized cosmic Chaplygin gas (GCCG) model. The striking factor
of the model, which made it unique, was that it can be made stable
and free from unphysical behaviours even when the vacuum fluid
enters the phantom era. In the previous studies related to DE
corresponding to phantom era it was seen that Big-Rip was the
final destiny, since the time derivative of scale-factor blows to
infinity in finite time. For the first time P. F. Gonz´alez-Diaz
introducing the GCCG model showed that Big Rip singularity can
easily be avoided. Hence in such models there is no requirement
for evaporation of black hole to zero mass.

The equation of state of Generalized cosmic Chaplygin gas is,
\begin{equation}\label{1}
p=-\rho^{-\alpha}\left[C+\left\{\rho^{1+\alpha}-C\right\}^{-\omega}\right]
\end{equation}
where $C=\frac{A}{1+\omega}-1$, with A being a constant that can
take on both positive and negative values, and $-L<\omega<0$, $L$
being a positive definite constant, which can take on values
larger than unity.

As numerous DE models began appearing in the scene, it was of
utmost necessity to devise a method that would both qualitatively
and quantitatively discriminate between various DE models. The
statefinder parameters \cite{sahni} were introduced in this
context by Sahni et al in 2003. These parameters are able to
discriminate between different DE in a model independent manner.
The diagnostic pair is constructed by the scale factor $a(t)$ and
its derivatives as follows:
\begin{equation}\label{1.1}
r\equiv\frac{\stackrel{...}a}{aH^3}~~~~~~~~~~s\equiv\frac{r-1}{3\left(q-\frac{1}{2}\right)}
\end{equation}
where $a$ is the scale factor of the universe, $H$ is the Hubble
parameter, a dot ($.$) denotes differentiation with respect to the
cosmic time $t$ and $q$ is the deceleration parameter given by,
\begin{equation}\label{1.2}
q=-\frac{\stackrel{..}a}{aH^{2}}
\end{equation}
In fact for $\Lambda$CDM model this pair take the values $r=1,
s=0$. For any DE model the trajectory in the $r-s$ plane is
obtained and the distance of this trajectory from the point
($1,0$) is noted. This produces the required discrimination.

In section 2, all discussions are valid in general for open,
closed and flat model of universe, i.e., for $k=\pm 1, 0$. But in
section 3, only the simple case of spatially flat universe ($k=0$)
is considered. In fact it is confirmed from various CMB
experiments \cite{benoit}, that the flat model of universe follows
spontaneously from the early inflationary phase, thus justifying
our consideration.

In this work we have tried to describe the GCCG model from the
theoretical point of view of a field, by introducing a scalar
field $\phi$ and a self interacting potential $V(\phi)$ with the
effective Lagrangian
\begin{equation}\label{1.3}
{\cal L}=\frac{1}{2}\dot{\phi}^{2}-V(\phi)
\end{equation}

In \cite{gor} it is shown that the flat Friedmann model with
Chaplygin gas can be described equivalently in terms of a
homogeneous minimally coupled scalar field $\phi$, by fitting the
FRW equations for Chaplygin gas into Barrow's scheme
\cite{barrow}. The expressions for homogeneous scalar field
$\phi(t)$ and potential $V(\phi)$ describing the Chaplygin
cosmology was obtained by Kamenshchik et al \cite{gor, gorini1,
gorini2}. In \cite{deb2} the work was repeated considering MCG as
the DE model. In this work we have obtained the corresponding
expressions for GCCG model as a generalization of the previous
works on Chaplygin gas cosmology.

The paper is organized as follows: In section 2 the expressions
for $\phi(t)$ and $V(\phi)$ are obtained in terms of the scale
factor $a(t)$ and the corresponding plots are generated to show
how $V(\phi)$ varies as the scale factor varies. In section 3, the
evolution of the universe is studied in the \{$r, s$\} plane for
the entire physically realistic history of the universe and
compared to that of the other Chaplygin gas models, especially
MCG. In section 4, a complete graphical analysis is performed. In
section 5 a comparative study between MCG and GCCG is performed
and finally the paper ends with some concluding remarks in section
6.

\section{GENERALIZED COSMIC CHAPLYGIN GAS IN FRW COSMOLOGY}
The metric of a homogeneous and isotropic universe in the FRW
model is

\begin{equation}\label{2}
ds^{2}=dt^{2}-a^{2}(t)\left[\frac{dr^{2}}{1-kr^{2}}+r^{2}\left(d\theta^{2}+\sin^{2}\theta
d\phi^{2}\right)\right]
\end{equation}
where $a(t)$ is the scale factor and $k(=0, \pm 1)$ is the
curvature scalar.

\noindent

The Einstein field equations are
\begin{equation}\label{3}
\frac{\dot{a}^{2}}{a^{2}}+\frac{k}{a^{2}}=\frac{1}{3}\rho
\end{equation}

and

\begin{equation}\label{4}
\frac{\ddot{a}}{a}=-\frac{1}{6}\left(\rho+3p\right)
\end{equation}
where $\rho$ and $p$ are the energy density and the isotropic
pressure, respectively. Here we choose $8\pi G=c=1$.

\noindent

The energy conservation equation is
\begin{equation}\label{5}
\dot{\rho}+3\frac{\dot{a}}{a}\left(\rho+p\right)=0
\end{equation}

Using equation (\ref{1}) in equation (\ref{5}), we have the
solution of $\rho$ as

\begin{equation}\label{6}
\rho=\left[C+\left\{1+\frac{B}{a^{3\left(1+\alpha\right)\left(1+\omega\right)}}\right\}^{\frac{1}{1+\omega}}\right]^{\frac{1}{1+\alpha}}
\end{equation}
where $B$ is an arbitrary integration constant.

Using eqns. (\ref{3}, \ref{4} and \ref{6}) we get the following
differential equation,
\begin{equation}\label{7}
H^{2}=\frac{1}{3}\left[C+\left\{1+B\left(1+z\right)^{3(1+\alpha)(1+\omega)}\right\}\right]^{\frac{1}{1+\alpha}}-k\left(1+z\right)^{2}
\end{equation}

where $z$ is the redshift.

Now for small values of scale factor $a(t)$, we have,
\begin{equation}\label{8}
\rho\simeq\frac{B^{\frac{1}{\left(1+\omega\right)\left(1+\alpha\right)}}}{a^{3}}
\end{equation}
which is a very large value and corresponds to the universe
dominated by an equation of state, $p=-\frac{1}{\rho^{Z}}$, where
$Z=\alpha\left(1+\omega\right)+\omega$ is a constant.

For large values of scale factor $a(t)$, we have,
\begin{equation}\label{9}
\rho\simeq\left(C+1\right)^{\frac{1}{1+\alpha}}
\end{equation}
For this value of $\rho$, the value of $p$ is found to be,
\begin{equation}\label{10}
p\simeq-\left(C+1\right)^{\frac{1}{1+\alpha}}=-\rho
\end{equation}
which corresponds to an otherwise empty universe with a
cosmological constant $\left(C+1\right)^{\frac{1}{1+\alpha}}$.

We know that an accelerating universe should have a negative
deceleration parameter, $q$. So it is clear from the expression of
deceleration parameter $(q=-\frac{\ddot{a}}{aH^{2}})$ that,
$\ddot{a}>0$. Some straightforward algebra using equations
(\ref{4} and \ref{6}), gives,
\begin{equation}
\left(1+\frac{B}{a^{3(1+\omega)(1+\alpha)}}\right)\left(1-2CB^{-\frac{1}{1+\omega}}a^{3(1+\alpha)}\right)-3<0
\end{equation}
i.e.,

$X\left(1-2CX^{-\frac{1}{1+\omega}}\right)<3$

where
$X=1+\frac{B}{a^{3\left(1+\omega\right)\left(1+\alpha\right)}}$.
Here the transition from decelerating to the accelerating universe
occurs at $a=\left(\frac{8C^{3}-1}{B}\right)^{\frac{1}{6}}$. So it
is quite clear from eqn.(15) that for large values of $a$, the
universe is accelerating whereas for smaller values of $a$, the
universe is decelerating. Hence in order to be consistent with the
cosmological observations we should consider large values of $a$
for our analysis. Considering the sub-leading terms in eqn.
(\ref{6}) we obtain for large values of scale factor, the
following expressions for energy density and pressure:

\begin{equation}\label{12}
\rho\simeq
C^{\frac{1}{1+\alpha}}+\frac{\left(Y-C\right)}{C^{\frac{\alpha}{1+\alpha}}\left(1+\alpha\right)}
\end{equation}

and\\

\begin{equation}
p\simeq-\left[C^{\frac{1}{1+\alpha}}+\frac{\left(Y-C\right)}{C^{\frac{\alpha}{1+\alpha}}\left(1+\alpha\right)}\right]
\end{equation}

where
$Y=C+\{1+\frac{B}{a^{3\left(1+\omega\right)\left(1+\alpha\right)}}\}^{\frac{1}{1+\omega}}$

Now we consider this energy density and pressure corresponding to
a scalar field $\phi$ having a self-interacting potential
$V(\phi)$. the Lagrangian of the scalar field is given by,
\begin{equation}
L_{\phi}=\frac{1}{2}\dot{\phi}^{2}-V(\phi)
\end{equation}
The expressions for energy density $\rho_{\phi}$ and pressure
$p_{\phi}$ for the scalar field are as follows,

\begin{equation}
\rho_{\phi}=\frac{1}{2}\dot{\phi}^{2}+V(\phi)=\rho=\left[C+\left\{1+\frac{B}{a^{3\left(1+\alpha\right)\left(1+\omega\right)}}\right\}^{\frac{1}{1+\omega}}\right]^{\frac{1}{1+\alpha}}=Y^{\frac{1}{1+\alpha}}
\end{equation}

and\\

$$p_{\phi}=\frac{1}{2}\dot{\phi}^{2}-V(\phi)=-\left[C+\left\{1+\frac{B}{a^{3\left(1+\alpha\right)\left(1+\omega\right)}}\right\}^{\frac{1}{1+\omega}}\right]^{-\frac{\alpha}{1+\alpha}}\left[C+\left(1+\frac{B}{a^{3\left(1+\alpha\right)\left(1+\omega\right)}}\right)^{-\frac{\omega}{1+\omega}}\right]$$
\begin{equation}
=-Y^{-\frac{\alpha}{1+\alpha}}\left[C+\left(Y-C\right)^{-\omega}\right]
\end{equation}

For a flat universe, i.e., $k=0$, we have,

$$\dot{\phi}^{2}=\left[C+\left\{1+\frac{B}{a^{3\left(1+\alpha\right)\left(1+\omega\right)}}\right\}^{\frac{1}{1+\omega}}\right]^{\frac{1}{1+\alpha}}\left[1-\left\{C+\left\{1+\frac{B}{a^{3\left(1+\alpha\right)\left(1+\omega\right)}}\right\}^{-\frac{\omega}{1+\omega}}\right\}\times\right.$$
\begin{equation}\label{phi}
\left.\left\{C+\left\{1+\frac{B}{a^{3\left(1+\alpha\right)\left(1+\omega\right)}}\right\}^{\frac{1}{1+\omega}}\right\}^{-1}\right]=Y^{\frac{1}{1+\alpha}}\left[1-\frac{1}{Y\left\{C+\left(Y-C\right)^{-\omega}\right\}}\right]
\end{equation}

and\\

$$V(\phi)=\frac{1}{2}\left[C+\left\{1+\frac{B}{a^{3\left(1+\alpha\right)\left(1+\omega\right)}}\right\}^{\frac{1}{1+\omega}}\right]^{\frac{1}{1+\alpha}}\left[1+\left\{C+\left\{1+\frac{B}{a^{3\left(1+\alpha\right)\left(1+\omega\right)}}\right\}^{-\frac{\omega}{1+\omega}}\right\}\times\right.$$
\begin{equation}
\left.\left\{C+\left\{1+\frac{B}{a^{3\left(1+\alpha\right)\left(1+\omega\right)}}\right\}^{\frac{1}{1+\omega}}\right\}^{-1}\right]=\frac{Y^{\frac{1}{1+\alpha}}}{2}\left[1+\frac{1}{Y\left\{C+\left(Y-C\right)^{-\omega}\right\}}\right]=Y^{\frac{1}{1+\alpha}}-\frac{1}{2}\dot{\phi}^{2}
\end{equation}

From the equation (\ref{phi}), we have the following relation
between scale factor $a(t)$ and scalar field $\phi$,
\begin{equation}
\phi=C'+\int_{1}^{t}\sqrt{\left\{C+\left(1+Bav^{-3\left(1+w\right)\left(1+\alpha\right)}\right)^{\frac{1}{1+w}}\right\}^\frac{1}{1+\alpha}\left(1-\frac{C+\left(1+Bav^{-3\left(1+w\right)\left(1+\alpha\right)}\right)^{\frac{-w}{1+w}}}{C+\left(1+Bav^{-3\left(1+w\right)\left(1+\alpha\right)}\right)^{\frac{1}{1+w}}}\right)}~~
dv
\end{equation}
where $C'$ is the integration constant.

Now
$2V(\phi)=\rho-p=\rho-\frac{C+\left(\rho^{1+\alpha}-C\right)^{-\omega}}{\rho^{\alpha}}$.\\

\noindent

{\bf Case I:}\\
When ~~~~$\rho\rightarrow\infty$,~~ i.e., ~~$a\rightarrow
0$,~~~~~~~~~~$V(\phi)\rightarrow\infty$

\noindent

{\bf Case II:} \\
When
$a\rightarrow\infty$,~~~~~~~~~~$V(\phi)\rightarrow\left(C+1\right)^{\frac{1}{1+\alpha}}$\\
For example,\\
When ~~$C=1$ ~~and~~ $\alpha=1$,
~~~~$V(\phi)\rightarrow\sqrt{2}$~~~ as ~~$a\rightarrow\infty$\\
When ~~$C=-1$ ~~~~~~~~~~~~~~~~~~~~~$V(\phi)\rightarrow 0$~~~~~~as
$a\rightarrow \infty$\\
When ~~$C=0$~~~~~~~~~~~~~~~~~~~~~~~~$V(\phi)\rightarrow 1$~~~~~~as $a\rightarrow \infty$\\

\begin{figure}

\includegraphics[height=2in]{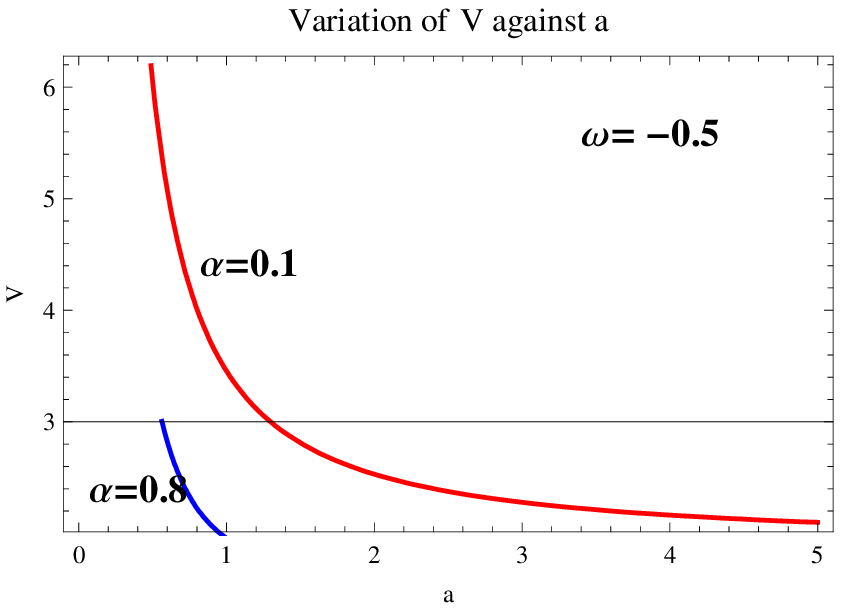}~~~~~~~~~~~~~~~~~~~~~\includegraphics[height=2in]{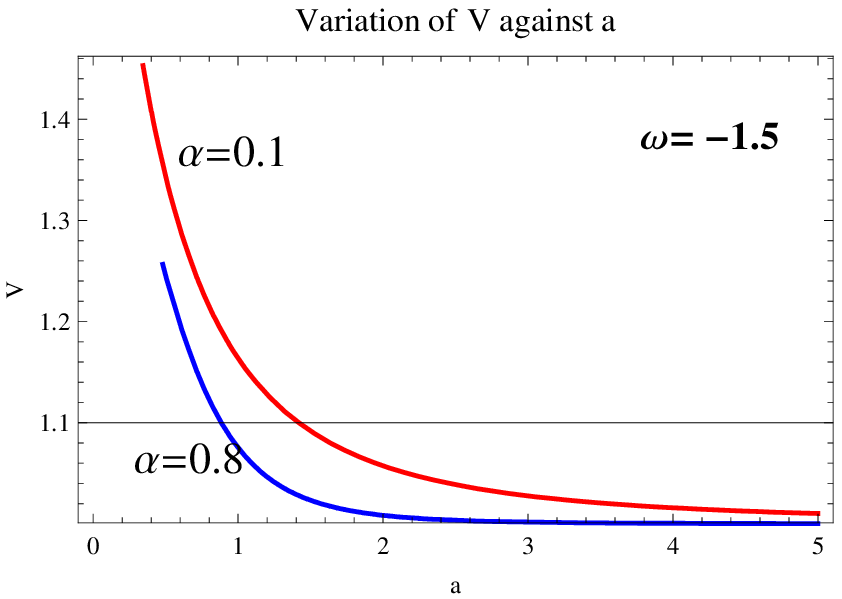}~~~\\
\vspace{1mm}
~~~~~~~~~~~~~Fig.1~~~~~~~~~~~~~~~~~~~~~~~~~~~~~~~~~~~~~~~~~~~~~~~~~~~~~~~~~~~~~~~~~~~~~~~~Fig.2~~~~~~~~~~~~~~~~~~~~\\

\vspace{3mm}

\textit{Figs.1 : The potential is plotted against time for different values of $\alpha$. The values of $\alpha$
used are $0.1$ and $0.8$. The value of $\omega$ is taken as $-0.5$.}\\

\textit{Fig.2 : The potential is plotted against time for different values of $\alpha$. The values of $\alpha$
used are $0.1$ and $0.8$. The value of $\omega$ is taken as $-1.5$.}\\

\end{figure}

\vspace{1mm}

\section{GEOMETRICAL DIAGNOSTIC USING STATEFINDER PARAMETERS}

\noindent

The statefinder parameters are defined as follows,
\begin{equation}\label{state}
r\equiv\frac{\stackrel{...}a}{aH^3}~~~~~~~~~~s\equiv\frac{r-1}{3(q-1/2)}
\end{equation}
where $H$ is the Hubble parameter and
$q=-\frac{a\ddot{a}}{\dot{a}^{2}}$ is the deceleration parameter.
Trajectories in the $r-s$ plane characterize different
cosmological models. The $\Lambda$CDM model corresponds to the
fixed point $\left\{1,0\right\}$ in the $r-s$ plane. The
deviations of the trajectories of different cosmological models
from this point in the $r-s$ plane gives an effective comparison
of the model from the standard model. So we proceed towards the
statefinder analysis of the present model, GCCG.\\

\noindent

For Friedmann model with a flat universe we have
\begin{equation}
H^{2}=\frac{\dot{a}^{2}}{a^{2}}=\frac{1}{3}\rho
\end{equation}
and\\
\begin{equation}
q=-\frac{\ddot{a}}{aH^{2}}=\frac{1}{2}+\frac{3}{2}\frac{p}{\rho}
\end{equation}
From the equation (\ref{state}), we have,
\begin{equation}\label{r}
r=1+\frac{9}{2}\left(1+\frac{p}{\rho}\right)\frac{\partial
p}{\partial
\rho}~~~,~~~~~~~~~s=\left(1+\frac{\rho}{p}\right)\frac{\partial
p}{\partial \rho}
\end{equation}
We get the ratio between $p$ and $\rho$ as,
\begin{equation}\label{p}
\frac{p}{\rho}=\frac{2\left(r-1\right)}{9s}
\end{equation}
For GCCG the velocity of sound can be given as,
\begin{equation}\label{v}
v_{s}^{2}=\frac{\partial p}{\partial \rho}=-\frac{\alpha
p}{\rho}+\frac{\omega
\left(1+\alpha\right)}{\left(\rho^{1+\alpha}-C\right)^{\omega+1}}
\end{equation}

From equations (\ref{r}), (\ref{p}) and (\ref{v}), we get the
following relation between $r$ and $s$
\begin{equation}\label{rs}
18s^{2}\left(9s\right)^{\frac{1}{\omega}}\left(r-1\right)+\left(9s\right)^{1+\frac{1}{\omega}}2\alpha\left(r-1\right)+4\alpha\left(9s\right)^{\frac{1}{\omega}}\left(r-1\right)^{2}=\omega\left(1+\alpha\right)\left(2-2r-9Cs\right)^{1+\frac{1}{\omega}}\left(9s+2r-2\right)
\end{equation}

\noindent

Now we consider three different cases:

{\bf Case 1: When $\omega=-\frac{1}{2}$}\\
For this case equation (\ref{rs}) becomes
\begin{equation}
2\left[18s^{2}\left(r-1\right)+18\alpha\left(r-1\right)s+4\alpha\left(r-1\right)^{2}\right]\left(2-2r-9Cs\right)+81s^{2}\left(1+\alpha\right)\left(9s+2r-2\right)=0
\end{equation}
From the above equation it is quite clear that there is only one
asymptote parallel to the $s$-axis, namely
$r=1+\frac{9\left(1+\alpha\right)}{4C}$. The asymptote intersects
the curve at only one point
$$\left(1+\frac{9\left(1+\alpha\right)}{4C}~,~~-\frac{A_{1}}{6B_{1}}-\frac{\left(-A_{1}^{2}+12B_{1}C_{1}\right)}{32^{2/3}B_{1}D_{1}+\sqrt{\left(D_{1}^{2}+4\left(-A_{1}^{2}+12B_{1}C_{1}\right)^{3}\right)^{1/3}}}+\frac{1}{62^{1/3}B_{1}}\left\{D_{1}\right.\right.$$
$$\left.\left.+\sqrt{\left(D_{1}^{2}+4\left(-A_{1}^{2}+12B_{1}C_{1}\right)^{3}\right)}\right\}^{1/3}\right)$$
where,\\
$A_{1}=1-18C+2\alpha-18C\alpha+4C^{2}\alpha+\alpha^{2}$\\
$B_{1}=\left(-17C^{2}+C^{2}\alpha\right)$\\
$C_{1}=\alpha+C^{2}\alpha+\alpha^{2}+C^{2}\alpha^{2}$\\
$D_{1}=-2+108C-1944C^{2}+11664C^{3}-12\alpha+540C\alpha-8412C^{2}\alpha+46872C^{3}\alpha-39600C^{4}\alpha+11016C^{5}\alpha-30\alpha^{2}+1080C\alpha^{2}-13560C^{2}\alpha^{2}+58968C^{3}\alpha^{2}-78648C^{4}\alpha^{2}+23112C^{5}\alpha^{2}-2448C^{6}\alpha^{2}-40\alpha^{3}+1080C\alpha^{3}-9648C^{2}\alpha^{3}+23976C^{3}\alpha^{3}-35976C^{4}\alpha^{3}+11448C^{5}\alpha^{3}-2432C^{6}\alpha^{3}-30\alpha^{4}+540C\alpha^{4}-2544C^{2}\alpha^{4}+216C^{3}\alpha^{4}+3000C^{4}\alpha^{4}-648C^{5}\alpha^{4}+144C^{6}\alpha^{4}-12\alpha^{5}+108C\alpha^{5}+12C^{2}\alpha^{5}-72C^{4}\alpha^{5}-2\alpha^{6}$

\vspace{4mm}

{\bf Case 2: When $\omega=-1$}\\

For this case equation (\ref{rs}) reads
\begin{equation}
2\left(r-1\right)\left[s+\alpha+\frac{2\alpha\left(r-1\right)}{9s}\right]+\left(1+\alpha\right)\left(9s+2r-2\right)=0
\end{equation}
Here the only asymptote parallel to the $s$-axis is
$r=1-\frac{9\left(1+\alpha\right)}{2}$. Here the asymptote
intersects the curve at the point
$\left(1-\frac{9\left(1+\alpha\right)}{2}~,~\frac{\alpha\left(1+\alpha\right)}{2\alpha+1}\right)$

\vspace{4mm}

{\bf Case 3: When $\omega=-2$}\\

Here the equation (\ref{rs}) transform into
\begin{equation}
9\left(1+\alpha\right)^{2}s\left(2-2r-9Cs\right)\left(9s+2r-2\right)^{2}+\left[9s^{2}\left(r-1\right)+9\alpha\left(r-1\right)s+2\alpha\left(r-1\right)^{2}\right]^{2}=0
\end{equation}
In this case we have two asymptotes parallel to $s$-axis, namely,
$r=1\pm 9\sqrt{C}\left(1+\alpha\right)$. The asymptote $r=1+
9\sqrt{C}\left(1+\alpha\right)$ intersects the curve at the point
$$\left(1+9\sqrt{C}\left(1+\alpha\right),~~\frac{A_{2}}{6B_{2}}-\frac{\left(-A_{2}^{2}+24B_{2}C_{2}\right)}{\left[32^{2/3}B_{2}\left\{D_{2}+\sqrt{\left(D_{2}^{2}+4\left(-A_{2}^{2}+24B_{2}C_{2}\right)^{3}\right)}\right\}^{1/3}\right]}+\frac{1}{62^{1/3}B_{2}}\left[D_{2}+\right.\right.$$
$$\left.\left.\sqrt{\left\{D_{2}^{2}+4\left(-A_{2}^{2}+24B_{2}C_{2}\right)^{3}\right\}}\right]^{1/3}\right)$$
where,\\
$A_{2}=8\sqrt{C}+4C^{3/2}+16\sqrt{C}\alpha-4C\alpha+8C^{3/2}\alpha+7\sqrt{C}\alpha^{2}-4C\alpha^{2}+4C^{3/2}\alpha^{2}$\\
$B_{2}=1+2C+\alpha-\sqrt{C}\alpha+2C\alpha$\\
$C_{2}=2C+6C\alpha+5C\alpha^{2}+C\alpha^{3}$\\
$D_{2}=128C^{3/2}+1344C^{5/2}+384C^{7/2}-128C^{9/2}+768C^{3/2}\alpha-192C^{2}\alpha+8064C^{5/2}\alpha-192C^{3}\alpha+2304C^{7/2}\alpha+384C^{4}\alpha-768C^{9/2}\alpha+2016C^{3/2}\alpha^{2}-960C^{2}\alpha^{2}+20352C^{5/2}\alpha^{2}-960C^{3}\alpha^{2}+6624C^{7/2}\alpha^{2}+1920C^{4}\alpha^{2}-1920C^{9/2}\alpha^{2}+2944C^{3/2}\alpha^{3}-2160C^{2}\alpha^{3}+27648C^{5/2}\alpha^{3}-2848C^{3}\alpha^{3}+11136C^{7/2}\alpha^{3}+3840C^{4}\alpha^{3}-2560C^{9/2}\alpha^{3}+2520C^{3/2}\alpha^{4}-2640C^{2}\alpha^{4}+21672C^{5/2}\alpha^{4}-4704C^{3}\alpha^{4}+10944C^{7/2}\alpha^{4}+3840C^{4}\alpha^{4}-1920C^{9/2}\alpha^{4}+1200C^{3/2}\alpha^{5}-1728C^{2}\alpha^{5}+9552C^{5/2}\alpha^{5}-3744C^{3}\alpha^{5}+5760C^{7/2}\alpha^{5}+1920C^{4}\alpha^{5}-768C^{9/2}\alpha^{5}+250C^{3/2}\alpha^{6}-480C^{2}\alpha^{6}+1896C^{5/2}\alpha^{6}-1120C^{3}\alpha^{6}+1248C^{7/2}\alpha^{6}+384C^{4}\alpha^{6}-128C^{9/2}\alpha^{6}$

\begin{figure}

\includegraphics[height=2in]{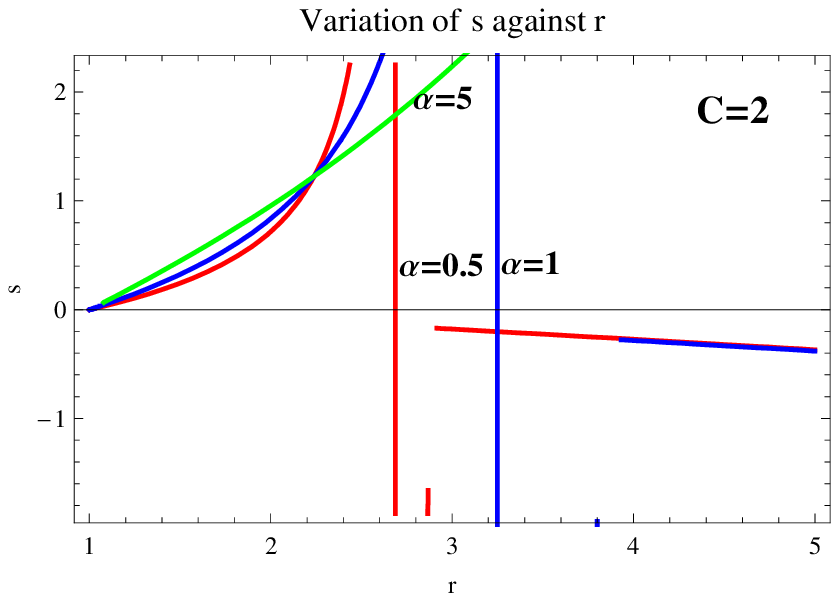}~~~~~~~~~~~~~~~~~~~~~~~~~~~~~~~~~\includegraphics[height=2in]{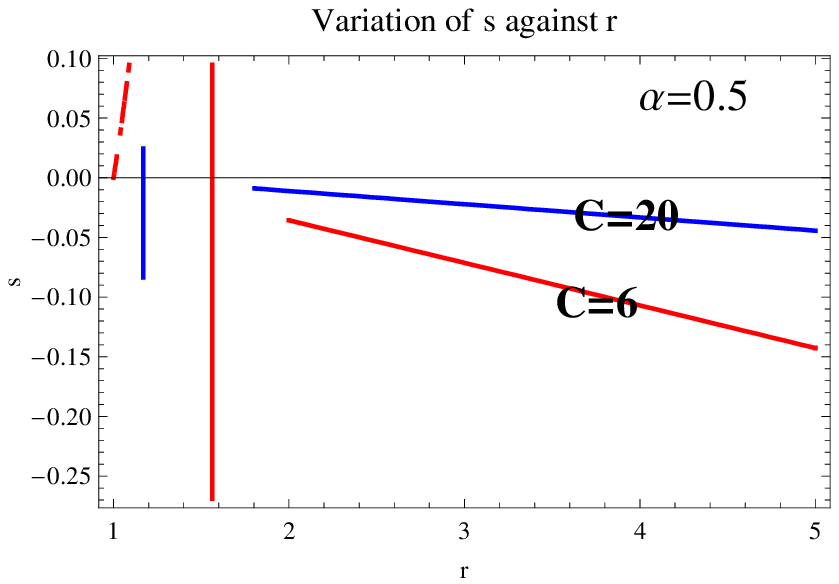}~~~~~~~~~\\
\vspace{1mm}
~~~~~~~~~~~~~~Fig. 3~~~~~~~~~~~~~~~~~~~~~~~~~~~~~~~~~~~~~~~~~~~~~~~~~~~~~~~~~~~~~~~~~~~~~~~~~~~~~~~~~Fig. 4~~~~~~~~~~\\

\vspace{3mm}

\includegraphics[height=2in]{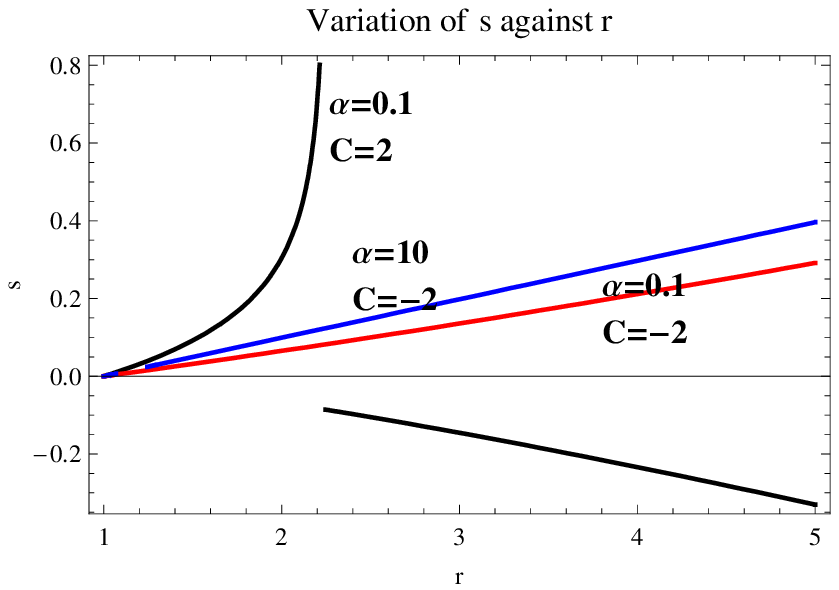}~~~~~\\
\vspace{1mm}
~~~~~~~~~~~~Fig. 5~~~~~~~~~~~\\

\vspace{3mm}

\textit{Fig.3 : The statefinder parameters are plotted against
each other for different values of $\alpha$. The values of $\alpha$ used are $0.5$, $1$ and $5$. The value of $C$ is taken as $2$.}\\

\textit{Fig.4 : The statefinder parameters are plotted agains each
other for different values of $C$. The values of $C$ used are $6$ and $20$. The value of $\alpha$ is taken as $0.5$.}\\

\textit{Fig.5 : The statefinder parameters are plotted against
each other for different values of $C$ and $\alpha$. The values
are furnished in the figure.}\\

\end{figure}

\vspace{1mm}

\section{GRAPHICAL ANALYSIS}
In the figures 1 and 2, the scalar field potential $V(\phi)$ is
plotted against time for different values of the parameters
involved. Two different plots are generated for different values
of $\omega$. It is evident that the scalar field potential
decreases as the scale factor increases. In an accelerating
universe the scale factor increases gradually with time. {\bf So
it is clear that the potential decreases with time.} Plots with
different values of $\alpha$ shows that the potential assumes
larger values for smaller $\alpha$. On the contrary a comparison
between figs. 1 and 2 indicate that {\bf with a decrease in the
value of $\omega$, there is corresponding decrease in the value of
potential.}

In figure 3, the statefinder parameters $r$ and $s$ are plotted
against each other for three different values of $\alpha$. It is
seen that for $r=1$, all the three curves tend towards $s=0$, thus
producing a correspondence with the $\Lambda$CDM model. With an
increase in the value of $\alpha$, there is a gradual shift of the
trajectories towards the right hand side of the plot, thus showing
an increasing $r$ tendency. In this plot $C=2$ is kept constant
throughout.

In Figure 4, the r-s trajectories are obtained for two different
values of $C$, keeping $\alpha$ fixed. It is seen that for $C=6$
(red curve), $s=0$ more or less corresponds with $r=1$, whereas
for $C=20$ (blue curve), $s=0$ does not exactly correspond to
$r=1$, thus showing a greater deviation. {\bf Thus it can be
concluded that as the value of $C$ increases, there is greater
deviation of the GCCG model from the standard $\Lambda$CDM model.}
Since $C$ can take both positive and negative values, r-s plots
for negative values of $C$ should be obtained and studied. This is
accomplished in fig.5. Three different values of $\alpha$ are
considered and the plots are generated.

\section{A COMPARATIVE STUDY BETWEEN THE ROLE OF GCCG AND MCG IN ACCELERATING UNIVERSE}
In \cite{deb} the authors studied the role of MCG in the
accelerating universe. In this section we undergo a comparative
study between this study and the study in \cite{deb}. We revisit
the plots for the scalar field potential and statefinder
parameters for MCG as witnessed in \cite{deb} in the figs. 6, 7,
8, 9 and 10. These plots when compared to the corresponding plots
of GCCG exhibits a few similarities and dissimilarities. We study
them separately below.

\begin{figure}

\includegraphics[height=2in]{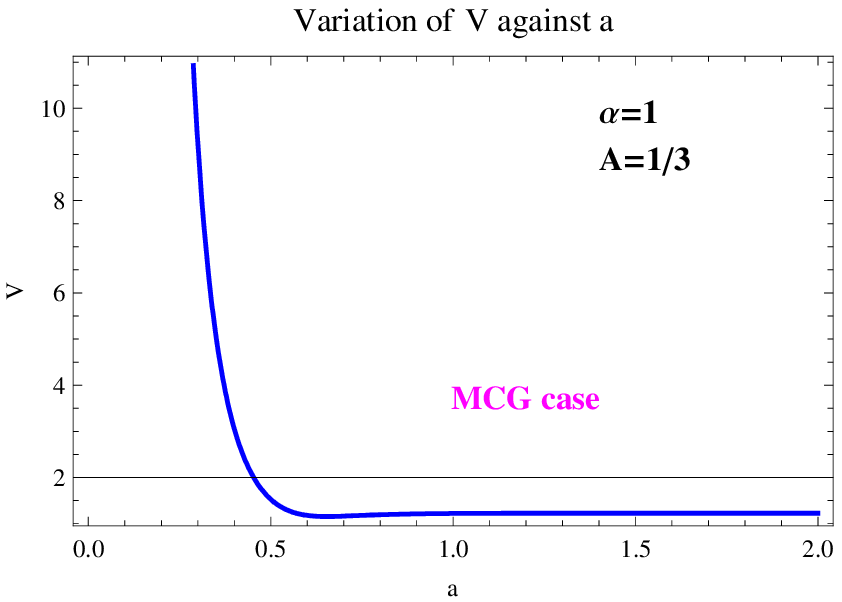}~~~~~~~~~~~~~~~~~~~~~~~~~~~~~~~~~\includegraphics[height=2in]{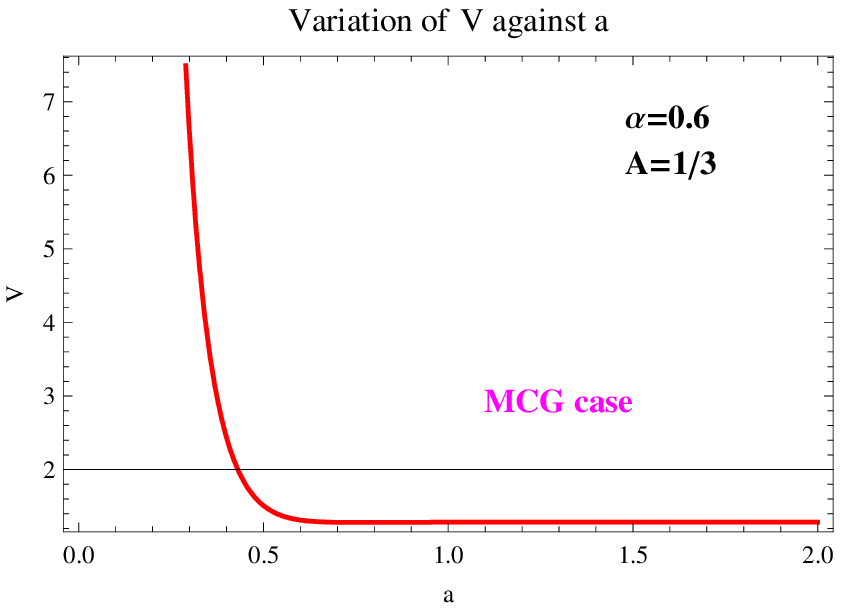}~~~~~~~~~\\
\vspace{1mm}
~~~~~~~~~~~~~~Fig. 6~~~~~~~~~~~~~~~~~~~~~~~~~~~~~~~~~~~~~~~~~~~~~~~~~~~~~~~~~~~~~~~~~~~~~~~~~~~~~~~~~Fig. 7~~~~~~~~~~\\

\vspace{2mm}

\textit{Fig.6 : The field potential is plotted against the scale factor in case of Modified Chaplygin gas.
The values of $\alpha$ and $A$ are respectively $1$ and $1/3$.}\\

\textit{Fig.7 : The field potential is plotted against the scale
factor in case of Modified Chaplygin gas. The values of $\alpha$ and $A$ are respectively $0.6$ and $1/3$.}\\

\vspace{2mm}

\includegraphics[height=2in]{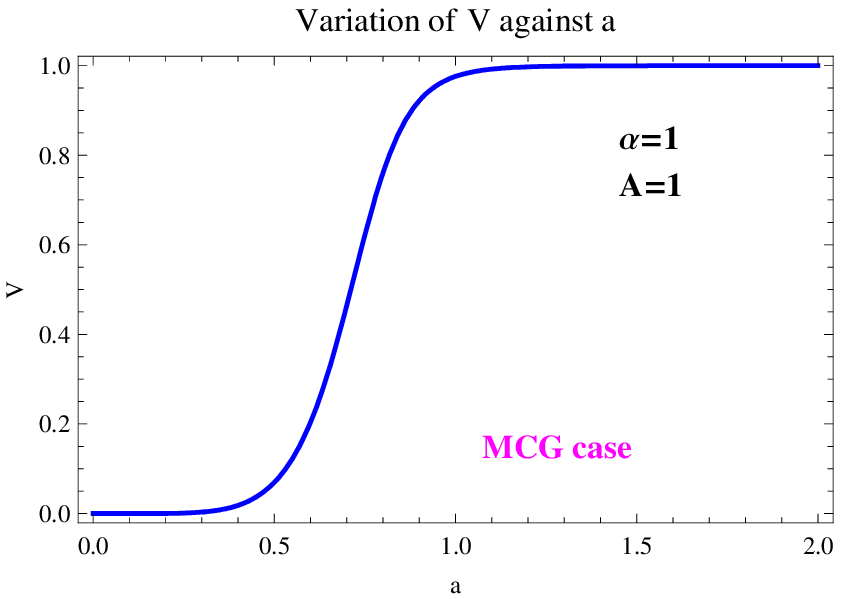}~~~~~~~~~~~~~~~~~~~~~~~~~~~~~~~~~\includegraphics[height=2in]{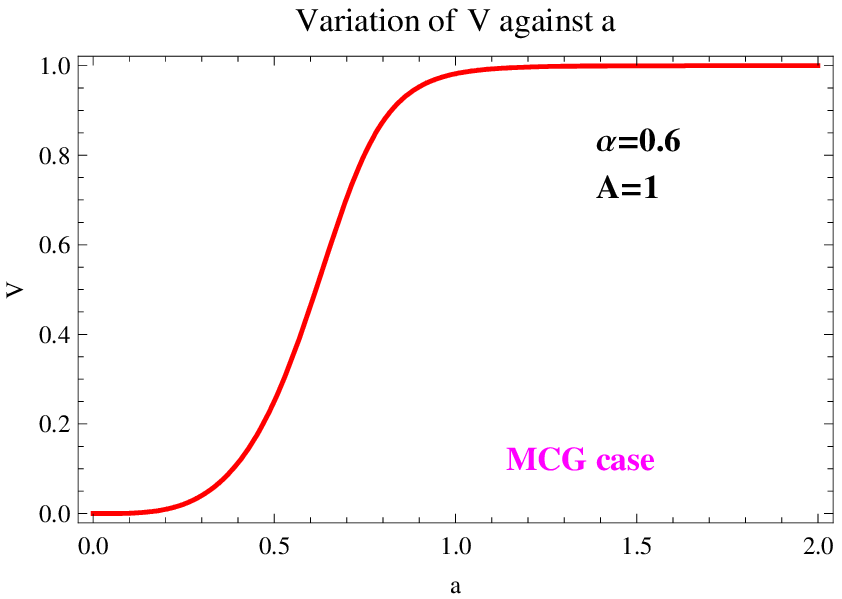}~~~~~~~~~\\
\vspace{1mm}
~~~~~~~~~~~~~~Fig. 8~~~~~~~~~~~~~~~~~~~~~~~~~~~~~~~~~~~~~~~~~~~~~~~~~~~~~~~~~~~~~~~~~~~~~~~~~~~~~~~~~Fig. 9~~~~~~~~~~\\

\vspace{3mm}

\textit{Fig.8 : The field potential is plotted against the scale
factor in case of Modified Chaplygin gas. The values of $\alpha$ and $A$ are respectively $1$ and $1$.}\\

\textit{Fig.9 : The field potential is plotted against the scale
factor in case of Modified Chaplygin gas. The values of $\alpha$ and $A$ are respectively $0.6$ and $1$.}\\

\vspace{2mm}

\includegraphics[height=2in]{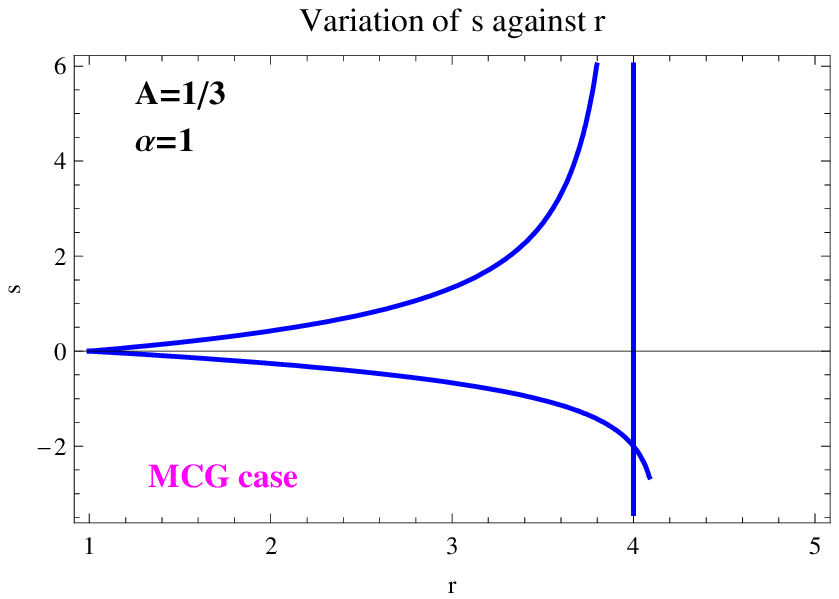}~~~~~\\
\vspace{1mm}
~~~~~~~~~~~~Fig. 10~~~~~~~~~~~\\

\vspace{3mm}

\textit{Fig.10 : The statefinder parameters $r$ and $s$ are
plotted against each other in case of Modified Chaplygin gas. The values of $\alpha$ and $A$ are respectively $1$ and $1/3$.}\\

\end{figure}

\subsection{Similarities}

We see that the plots 1 and 2 for GCCG almost show similar
characteristics as the plots 6 and 7 for MCG, taking $A=1/3$. In
these figures the potential decrease with increase in scale
factor.

\subsection{Differences}

1)  In the plots 8 and 9 for MCG, taking $A=1$, the curve behave
quite differently from the plots 1 and 2 for GCCG. In case of MCG
the potential increases steeply for a while before becoming
stagnant at a particular value of $V$. But in case of GCCG there
is a constant decrease in potential with increase in $a$, as seen
before.

\vspace{5mm}

2) Fig 10 shows the $r-s$ plot for MCG. When compared with the
$r-s$ plots for GCCG (figs. 3, 4 and 5), clear differences in
trajectories are observed. This helps us to discriminate between
the two DE models.

3) In fig. 10, the portion of the curve on the positive side of
$s$, which is physically admissible is only the values of $r$
greater than $\{1+\frac{9}{2}A\left(1+A\right)\}$ (for $A=1/3$, it
becomes equal to 3). Incidentally the portion of the curve between
$r=1$ and $r=\{1+\frac{9}{2}A\left(1+A\right)\}$ on the positive
side of $s$ is not admissible and that portion of the curve is not
shown in \cite{deb}. The reason behind this is that in this
restricted region constant $B$ becomes negative, which is
inadmissible. Therefore the curve on the positive side of $s$
starts from the radiation era and goes asymptotically to the dust
model, but the portion on the negative side of $s$ shows the
evolution right from the dust state ($s=-\infty$) to the
$\Lambda$CDM model ($s=0$). Hence the negative and the positive
region complement each other, and the total curve (both positive
$s$ and negative $s$ portions) represents the evolution of the
universe from the radiation era to the $\Lambda$CDM model.

But in case of GCCG we see that there is no such restriction on
any of the parameters $C$, $\omega$ and $\alpha$, since they do
not have any restriction on either magnitude or sign. So the
entire portion of the curves in figs. 3, 4 and 5 is physically
admissible and there is no restricted portion, as it is in case of
MCG. This is a striking difference! {\bf In fact it makes GCCG a
much less constrained form of dark energy.}

\section{CONCLUSION}
In this work we have considered Generalized cosmic Chaplygin gas
and tried to find the role played by it in an accelerating
universe described by FRW cosmology. To accomplish this we have
compared the GCCG model with a scalar field, and determined its
potential. Plots were generated for the determined potential.
Statefinder parameters were studied in detail in order to
discriminate the DE model from other DE models. We have undergone
a comparative study between the GCCG and MCG models as far as
their role in the accelerating universe is concerned. The
trajectories of the $r-s$ plots helps us to discriminate between
the two models. For $A=1$, it was found that MCG shows an increase
in the scalar potential in a certain region, as $a$ increases. But
in case of GCCG, irrespective of whatever value is assigned to the
parameters, the potential always show a downhill progress. It was
witnessed that unlike MCG there is no restricted region in the
$r-s$ plot for GCCG, thus making every portion of the curve
physically admissible as it violates the admissible values of the
B parameter of MCG. In case of GCCG there cannot be any such
inadmissible region in the r-s plot. This is possible because the
restrictions of the $C$, $\omega$ and $\alpha$ parameters of GCCG
is much less pronounced than those of the parameters of MCG. .
Hence GCCG is less constrained compared to MCG. This result is
important because it gives a more universal character to GCCG as a
dark energy. It is shown in \cite{Rudra1} that GCCG as a dark
energy is less effective than MCG and other dark energy models. In
the background of this result our current finding of the universal
nature of GCCG as a dark energy is quite unexpected! A dark energy
suffering from much less restriction imposed by its parameters can
be physically and mathematically more acceptable compared to those
dark energy models which work under a very small domain with a lot
of parameter constraints. It provides an independent outlook to
the model capable of adapting itself to any domain of cosmology.
Finally it should be stated that GCCG describes the universe from
radiation era to $\Lambda$CDM era without constraining its
parameters considerably.

\end{document}